\documentclass[letterpaper]{article} 
\usepackage{aaai25}  
\usepackage{times}  
\usepackage{helvet}  
\usepackage{courier}  
\usepackage[hyphens]{url}  
\usepackage{graphicx} 
\usepackage{natbib}  
\usepackage{caption} 
\usepackage{algorithm}
\usepackage{algorithmic}
\usepackage{subcaption}

\usepackage{newfloat}
\usepackage{listings}
\DeclareCaptionStyle{ruled}{labelfont=normalfont,labelsep=colon,strut=off} 
\floatstyle{ruled}
\newfloat{listing}{tb}{lst}{}
\floatname{listing}{Listing}
\title{AI-Empowered Human Research Integrating Brain Science and \\Social Sciences Insights}
\author{
    Feng Xiong\textsuperscript{\rm 1},
    Xinguo Yu\textsuperscript{\rm 1},
    Hon Wai Leong\textsuperscript{\rm 2}
}
\affiliations{
    \textsuperscript{\rm 1}Faculty of Artiﬁcial Intelligence in Education, Central China Normal University\\
    \textsuperscript{\rm 2}Department of Computer Science, National University of Singapore\\
    mountain@mails.ccnu.edu.cn,
    xgyu@ccnu.edu.cn, 
    leonghw@comp.nus.edu.sg
}

\usepackage{bibentry}

\begin{document}

\maketitle

\begin{abstract}
This paper explores the transformative role of artificial intelligence (AI) in enhancing scientific research, particularly in the fields of brain science and social sciences. We analyze the fundamental aspects of human research and argue that it is high time for researchers to transition to human-AI joint research. Building upon this foundation, we propose two innovative research paradigms of human-AI joint research: ``AI-Brain Science Research Paradigm'' and ``AI-Social Sciences Research Paradigm''. In these paradigms, we introduce three human-AI collaboration models: AI as a research tool (ART), AI as a research assistant (ARA), and AI as a research participant (ARP). Furthermore, we outline the methods for conducting human-AI joint research. This paper seeks to redefine the collaborative interactions between human researchers and AI system, setting the stage for future research directions and sparking innovation in this interdisciplinary field.
\end{abstract}

%

\section{Introduction}

As artificial intelligence (AI) continues to advance rapidly, it has become an integral component of daily life, particularly within scientific research. The ability of AI to process vast datasets, simulate complex systems, and generate meaningful insights has fundamentally revolutionized research paradigms. This impact is particularly pronounced in the fields of brain science and social sciences, where AI is transitioning from a tool to an active research collaborator. This indicates that AI is no longer a mere instrument but is increasingly becoming an autonomous entity actively participating in the research process. Simultaneously, the scope of research is expanding to include the study of AI's behavior and its impact on society. These transformations compel a reconsideration of traditional research frameworks, necessitating a shift from human-centered research to human-AI joint research, where AI plays an active and independent role in scientific inquiry.

Human-AI joint research refers to an innovative research paradigm in which AI acts as an autonomous entity, collaborating with human researchers in generating insights and solutions. In several fields, preliminary applications of AI in research assistance represent the nascent stages of this paradigm. In brain science, Krauss \shortcite{krauss2024understanding} illustrates how AI can aid in understanding the complexities of the brain, while Wang et al. \shortcite{wang2024brain} examine the interactions between AI and cognitive processes. Zhao et al. \shortcite{zhao2023brain} caution that the convergence of brain-inspired AI with artificial general intelligence (AGI) could bring about unprecedented changes. Adesso \shortcite{adesso2023towards} demonstrates the significant potential of AI in brain science, particularly in driving scientific discovery. In the social sciences, Bail \shortcite{bail2024can} argues that generative AI could fundamentally alter research methods and outcomes, while Xu et al. \shortcite{xu2024ai} explore the bidirectional interactions between AI and social sciences. Grossmann et al. \shortcite{grossmann2023ai} highlight AI's profound impact on social sciences research paradigms, and Pérez et al. \shortcite{perez2023serious} show how AI's integration with serious games can open new research avenues. Farooq et al. \shortcite{farooq2023ai} provide a comprehensive review showing rapid advances in AI's application in social sciences. Graziani et al. \shortcite{graziani2023global} promote interdisciplinary research by advocating for the global standardization of explainable AI terminology across technology and social sciences.

Despite AI's growing role in scientific research, a comprehensive framework for leveraging its full potential in collaborative research is still lacking. Achieving human-AI joint research requires focusing on three key areas: (1) AI system must continuously advance, acquiring higher cognitive and social interaction abilities, such as complex reasoning, empathy, and adaptive collaboration. These capabilities are critical for AI to effectively engage with human researchers. (2) It is necessary to study human cognitive and social interaction mechanisms from the perspectives of brain science and social sciences, as these abilities drive innovation in the human research process. (3) Developing effective human-AI collaboration models is essential to enhance joint research capabilities, fostering unprecedented levels of cooperation and innovation in scientific research.

To address the three key areas mentioned earlier, this paper first reviews the fundamental aspects of human research from the perspectives of brain science and social sciences. We examine how cognitive and emotional processes, studied within brain science, shape human behavior and decision-making in research. Additionally, we explore how social sciences analyze interactions between individuals and groups, providing insights into collaboration and knowledge sharing. By understanding these human-driven processes, researchers can design AI systems that more accurately simulate human cognition and social interaction. This development lays the foundation for AI's integration as a genuine collaborator in scientific research.

We then propose two innovative paradigms of human-AI joint research: ``AI-Brain Science Research Paradigm'' and ``AI-Social Sciences Research Paradigm''. In these paradigms, we identify three distinct modes of human-AI collaboration: AI as a research tool (ART), AI as a research assistant (ARA), and AI as a research participant (ARP). Furthermore, we outline methods for conducting human-AI joint research, including empirical studies and questionnaire-based surveys to evaluate the impact of AI on creative and critical thinking within the research process. These paradigms and methods provide a structured approach for integrating AI into scientific research as a full collaborator alongside human researchers.

The main contributions of this paper are as follows:
\begin{enumerate}
\item It reviews the foundational aspects of human cognition and social interaction in human research process, providing insights critical for developing human-AI joint research.
\item It proposes two innovative research paradigms and identifies three distinct human-AI collaboration models in human-AI joint research.
\item It introduces methods for conducting human-AI joint research, providing practical approaches for integrating AI as a full research partner in collaborative research.
\end{enumerate}

\section{Related Work}
This section reviews AI's integration into scientific work, examining its expanding role and the emerging challenges that accompany this evolution. It also covers the challenges of integrating AI with traditional discipline research field, emphasizing the need for new research paradigms. Finally, the section also discusses ethical concerns such as bias, privacy violations, and intellectual property, appealing the importance of establishing clear ethical guidelines for the future of human-AI joint research.
\subsection{Preliminary Application of AI in Scientific Research}
Fui-Hoon Nah et al. \shortcite{fui2023generative} note that generative AI, such as ChatGPT, is emerging as a significant force in research writing, though it also faces unprecedented challenges. Spillias et al. \shortcite{spillias2023human} demonstrate AI's potential in literature identification, while Han et al. \shortcite{han2024teams} explore collaborative strategies when teams leverage AI for creative design tasks. Collectively, these studies indicate that while AI's role in research is rapidly expanding, its full potential has yet to be realized. Fragiadakis et al. \shortcite{fragiadakis2024evaluating} further highlight that the interaction between AI and humans remains an area requiring in-depth study.

\subsection{Challenges of Integrating AI with Traditional Disciplines}
As AI deeply integrates with traditional disciplines, researchers face unprecedented challenges. Zhao et al. \shortcite{zhao2023brain} highlight that while brain-inspired AI has made progress in simulating human cognition, it struggles with fully replicate human intelligence such as reasoning and thinking. As artificial general intelligence (AGI) AGI advances, these challenges become more pronounced, revealing the limitations of traditional brain science methods in addressing AI's growing cognitive demands. Besides, Steyvers and Kumar \shortcite{steyvers2024three} further emphasize the complexities and uncertainties AI faces in decision-making processes. Traditional social sciences research methods often fall short in capturing the intricate dynamics of human-AI interactions, particularly in emotional and social contexts. These limitations point out the need for new frameworks to fully integrate AI in traditional discipline research.

Hardy et al. \shortcite{hardy2023large} underscore the limitations of large language models (LLMs) in comprehending and generating natural language. While LLMs perform well in standard language tasks, their deficiencies in handling emotional nuances indicate a need for further refinement. Similarly, Huang \shortcite{huang2024eight} identifies eight key challenges in the development of intelligence theories, including complexity, diversity, and dynamism. He notes the significant limitations of current brain science and AI research methods in addressing complex social behaviors and cross-cultural emotions, stressing the need for expanding and innovating these frameworks.

\subsection{Emergence of New Research Paradigms}
Bozkurt \shortcite{bozkurt2023generative} discusses the paradigm shifts triggered by generative AI in the field of education. He notes that traditional educational methods and cognitive theories are increasingly inadequate in meeting modern needs. Several studies \cite{fragiadakis2024evaluating, rezwana2023designing} systematically review and design various forms of human-AI collaboration, proposing new evaluation standards. This work further demonstrates the inadequacy of traditional research frameworks in addressing the multidimensional complexities of human-AI collaboration. Shen et al. \shortcite{shen2024towards} explore a systematic framework for bidirectional human-AI alignment, emphasizing the importance of considering the reciprocal effects of human-AI interaction rather than understanding AI's role solely from a technological perspective. 

In general, these studies suggest that while AI has shown tremendous potential in scientific research, the limitations of existing research paradigms have hindered its full utilization. To overcome these challenges and fully unleash AI's potential, the development of ``AI-Brain Science Research Paradigm'' and ``AI-Social Sciences Research Paradigm'' has become imperative. These new paradigms will not only address the shortcomings of current research but also offer new perspectives and pathways for future scientific inquiry.

\subsection{Ethical Considerations in Human-AI Joint Research }
 
The integration of AI into research raises several ethical concerns, particularly regarding behavioral restriction and data usage. B. Stahl and D. Eke \shortcite{stahl2024ethics} highlight the risks of bias in AI decision-making, especially when sensitive data is used for training. As AI acts as research participants, ensuring transparency and fairness in its actions is crucial. Wu et al. \shortcite{wu2024unveiling} emphasize the need for ethical frameworks to address bias in AI's actions, particularly when its decisions influence human research outcomes. 

Additionally, issues such as privacy violations and copyright infringement are significant concerns in AI-driven research. The use of large datasets often involves sensitive information, raising the risk of privacy breaches. Copyright issues can also arise when AI-generated content relies on pre-existing materials without proper attribution or permission. Establishing clear ethical guidelines to address these concerns is essential for developing human-AI joint research.

\section{Fundamentals of Human Research}
This part examines the fundamental aspects of human research processes that are critical for the human-AI joint research. As AI becomes an increasingly indispensable partner in scientific research, ensuring its behavior is reliable and well-suited for collaboration with humans is necessary. Therefore, it is crucial to understand the core human factors that drive research---cognition, emotion, and collaboration.

In the past few decades, extensive research has been conducted on information processing and social interaction within brain science and social sciences. Brain science aims to uncover the complexities of human brain function and behavior, with the ultimate goal of fully understanding how the brain is structured and operates. Besides, social sciences focus on understanding the interactions between individuals and groups, exploring how social dynamics shape human behavior and relationships in various contexts.

By reviewing these fundamental theoretical studies on human brain information processing and human-human social interaction, we can gain valuable insights that will guide us in exploring the emerging research field of ``human-AI joint research''.

\begin{figure*}[h!]
\centering
\includegraphics[width=0.9\textwidth]{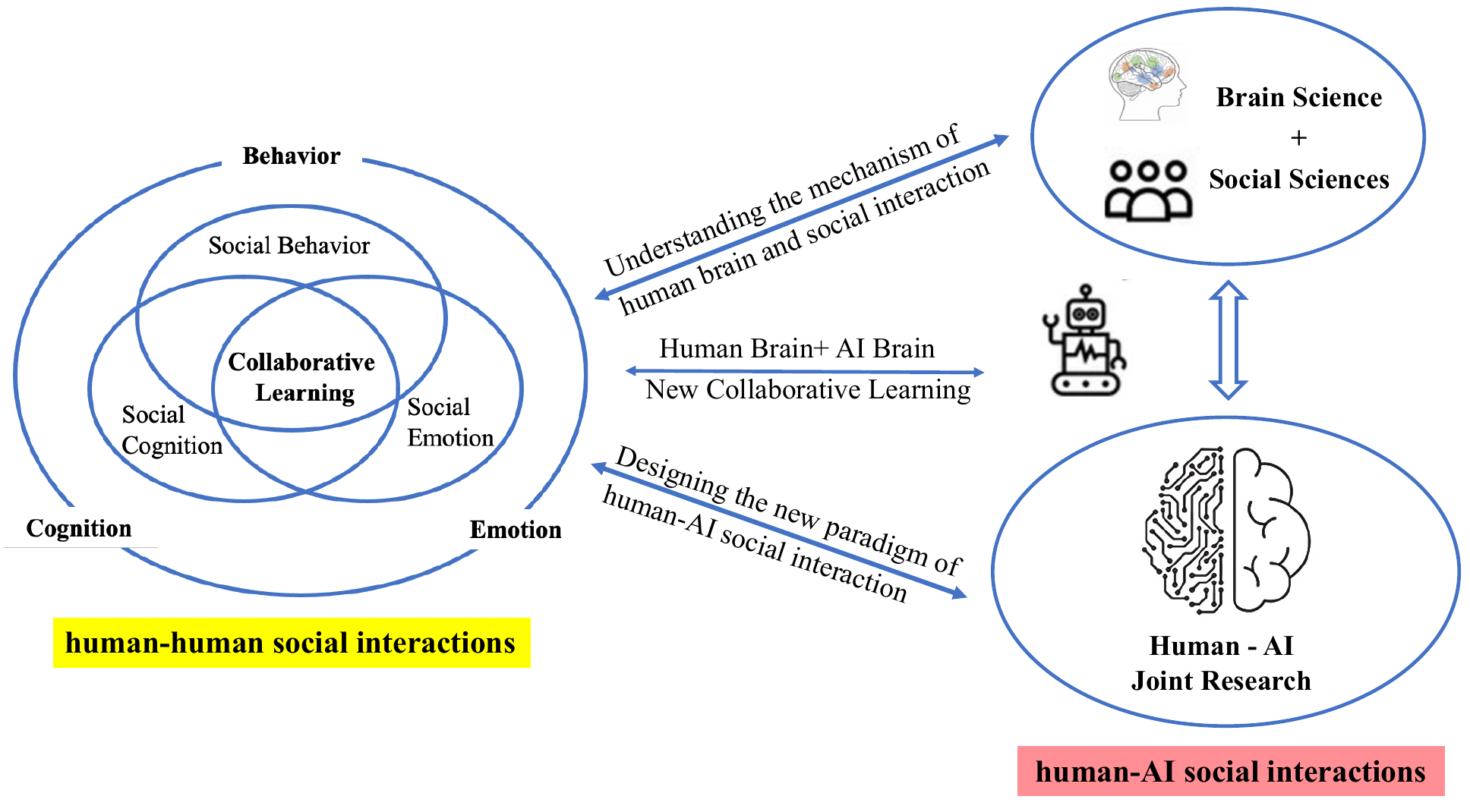} 
\caption{From human-human social interaction to human-AI social interaction. The left part describes the fundamentals of human collaboration. The right part represents the inclusion of AI in human interactions forms a new paradigm, where human and AI actively collaborate. Insights from brain science and social sciences will drive advancements in human-AI joint research.}
\label{fig:human2ai}
\end{figure*}

\subsection{Cognition and Emotion in Human Research}
Human research processes are driven by advanced cognitive and emotional capabilities, both of which are central to the field of brain science. In scientific research, cognitive functions allow individuals to absorb and process vast amounts of information, make decisions, solve problems, and innovate \cite{sommerville2020social, dehay2020evolution}. At the same time, emotions play a vital role in shaping motivation, collaboration, and communication, which are essential in research environments. Emotional responses influence how researchers interact with their peers, how they handle challenges, and how they interpret and respond to feedback during the collaborative process \cite{phelps2006emotion}. 

Traditionally, cognition and emotion have been viewed as separate domains. For centuries, scholars such as Plato have maintained that these two aspects of human experience were distinct \cite{scherer1984onthenature, lazarus1999cognition}. However, recent findings of brain science indicate that cognition and emotion are deeply intertwined, working together to shape behavior and decision-making \cite{pessoa2013cognitive, pessoa2008relationship}. This connection is especially significant in research, where cognitive functions like attention, memory, and thinking are strongly influenced by emotional states \cite{dolan2002emotion, dolcos2020neural}. 

By understanding the connection between cognition and emotion from the perspective of brain science, we can gain deeper insights into how researchers operate and collaborate. This knowledge forms the basis for developing advanced research AI and promoting human-AI joint research.

\subsection{Fundamentals of Human Collaborative Learning}
Collaborative learning is a process where individuals work together to generate ideas, share knowledge, and form connections within groups. This interaction is shaped by both cognitive and emotional processes \cite{kreijns2003identifying, goffman1983interaction}, making it an essential framework for understanding the dynamics of human collaboration. In the context of human-AI joint research, these theories provide foundations for studying how AI can collaborate with humans effectively.

Cognitive processes in collaborative learning involve the thinking and reasoning that individuals use to create shared knowledge \cite{kreijns2003identifying}. According to socio-cognitive theory, behavior is influenced by mental processes, emotional states, and the social environment \cite{usher2017social}. Individuals not only regulate their actions and adapt to various situations but also shape, and are shaped by, their environment. These cognitive exchanges form the intellectual foundation of collaboration, allowing groups to solve complex problems and innovate collectively.

Equally important are the socio-emotional processes involved in collaboration, which govern how individuals interact and express emotions within the group. These emotional bonds enhance the learning experience, fostering trust and openness that facilitate knowledge exchange \cite{janssen2012task}. Just as cognition drives knowledge acquisition, emotions and social interactions are influenced by perceptions and beliefs about others in the group \cite{van2006social}. For collaboration to be effective, both cognitive and emotional engagement are necessary. Collaborative dialogue, where individuals reflect on and share different perspectives, improves learning outcomes and promotes the generation of new ideas \cite{dillenbourg1999you, chi2014icap}.

\subsection{From Human Research to Human-AI Joint Research}
As previously described, in the exploration of human-AI joint research, brain science and social sciences play essential roles. As illustrated in Figure \ref{fig:human2ai}, brain science focuses on how cognition and emotion influence behavior, providing insights into the workings of the human brain. This understanding helps explore how AI can simulate and enhance human thought processes in research. On the other hand, social sciences study the interactions between individuals and groups. This provides the foundation for understanding interaction mechanisms, which is critical for examining AI's role and function in research. 

At the same time, traditional research methods in brain science and social sciences must adapt to accommodate the transition toward human-AI joint research. Now is a critical moment for this transformation. Advancing human-AI joint research will not only revitalize current research methods but also foster the development of more forward-looking research paradigms. This transition will lead to closer collaboration between AI and humans in the research process, enabling AI to evolve from a tool to a full research partner, ushering in a new era of scientific research.

\begin{figure*}[h!]
\centering
\begin{subfigure}{\linewidth}
    \centering
    \includegraphics[width=0.9\textwidth]{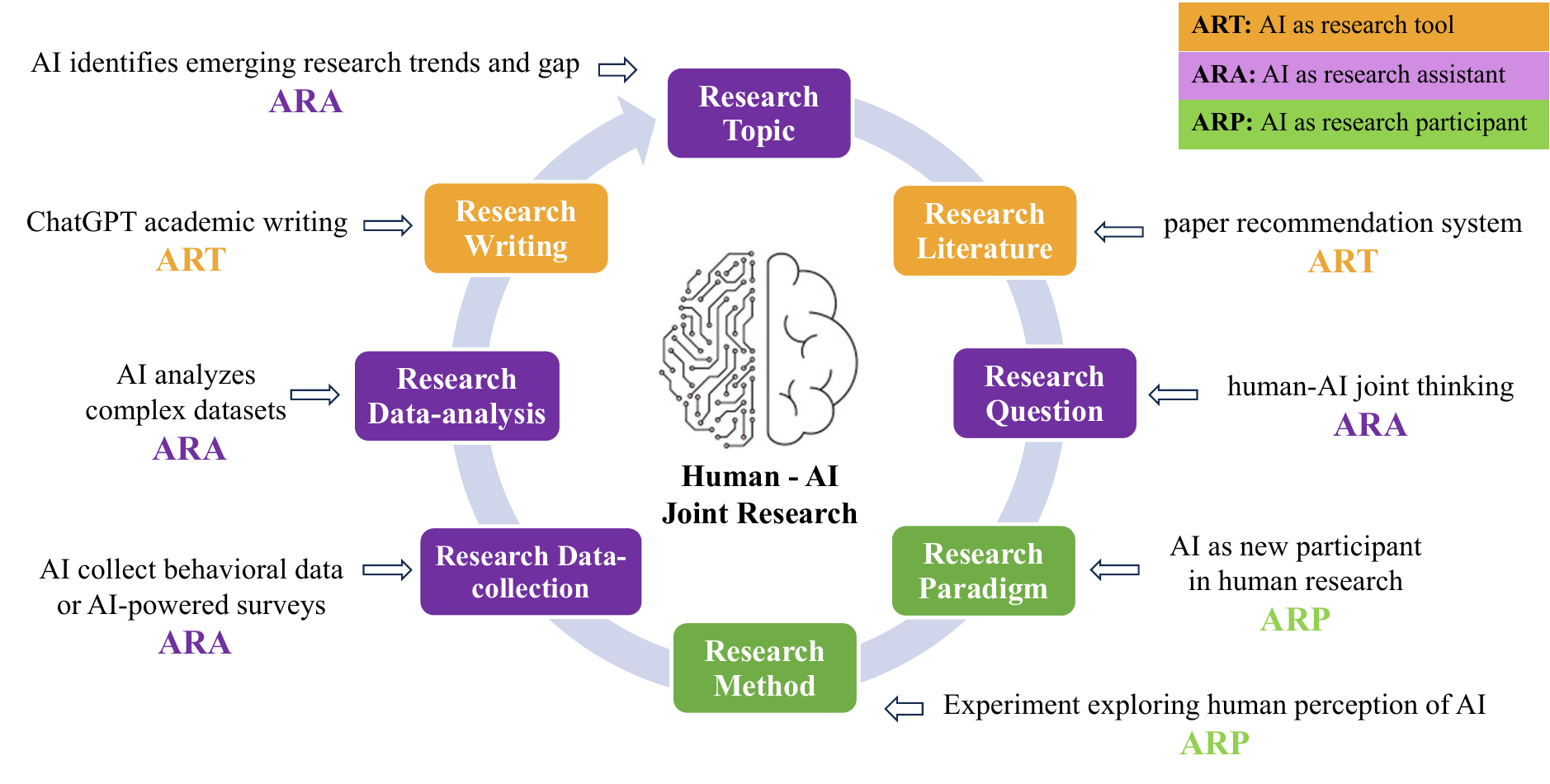}
    \caption{Three human-AI collaboration models in research process}
\end{subfigure}
\begin{subfigure}{\linewidth}
    \centering
    \includegraphics[width=0.85\textwidth]{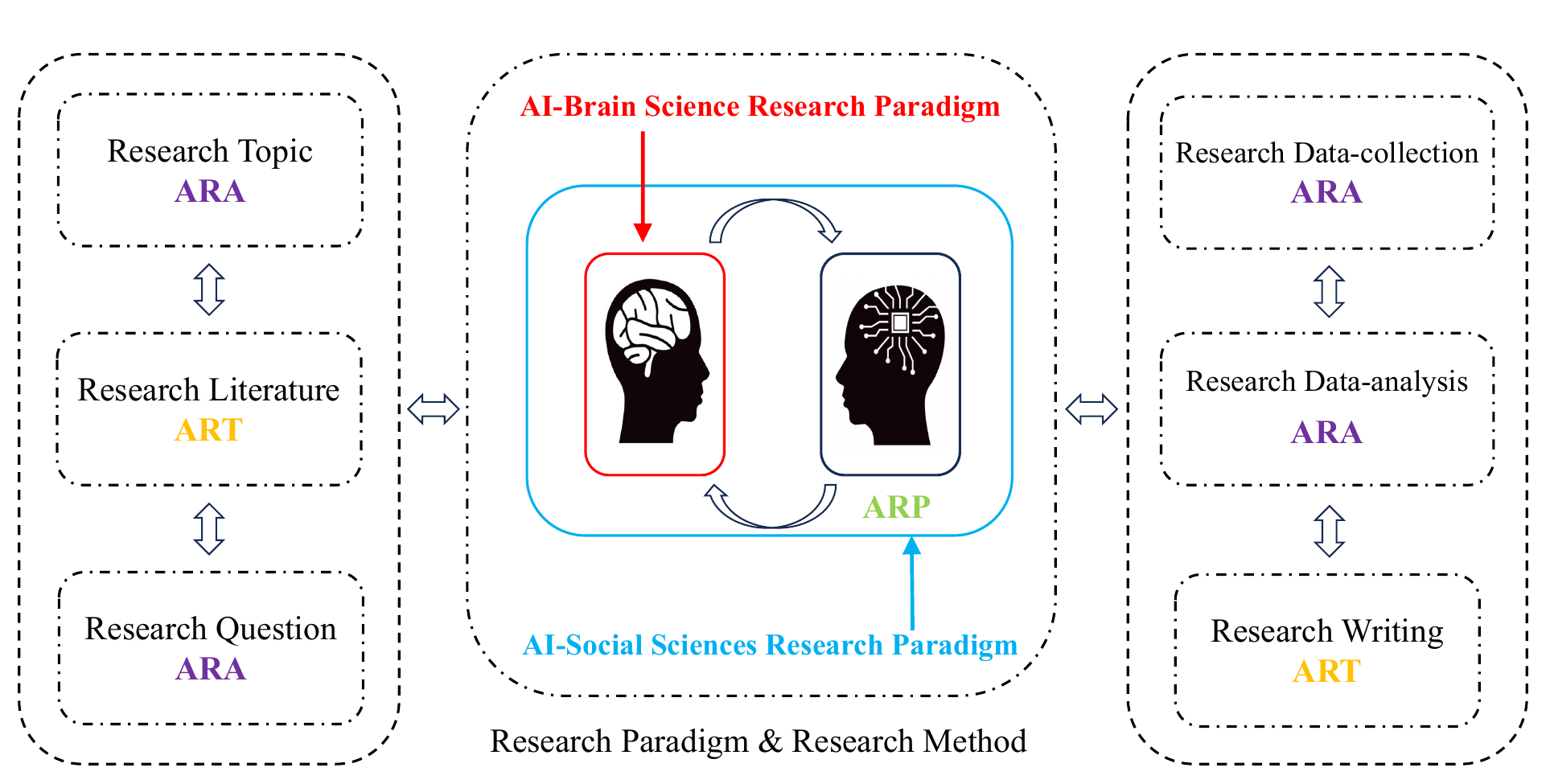}
    \caption{New research paradigms}
\end{subfigure}
\caption{Human-AI collaboration models and new research paradigms of human-AI joint research. (a) Three human-AI collaboration models in research process: AI as a research tool (ART), AI as a research assistant (ARA), and AI as a research participant (ARP). (b) New research paradigms emerging from AI's integration into brain science and social sciences: AI-Brain Science Research Paradigm and AI-Social Science Research Paradigm.}
\label{fig:AI-role}
\end{figure*}

\section{New Research Paradigms of Human-AI Joint Research}
In the human-AI joint research field, the combination of AI with traditional brain science and social sciences will produce two new research paradigms. The first research paradigm is based on the traditional research paradigm of brain science and considers the impact of AI on human brain cognition and emotional processing, resulting in a new research paradigm ``\textbf{AI-Brain Science Research Paradigm}''. The second research paradigm is based on the traditional research paradigm of social sciences and considers the impact of AI on the social interaction of individuals or groups, resulting in a new research paradigm ``\textbf{AI-Social Sciences Research Paradigm}''. As shown in Figure \ref{fig:AI-role}, in these two new research paradigms of human-AI joint research, there are three distinct modes of human-AI collaboration: (1) AI as a research tool (\textbf{ART}); (2) AI as a research assistant (\textbf{ARA}); (3) AI as a research participant (\textbf{ARP}).

\subsection{AI as Research Tool Empowers Human Research}
AI as a research tool (ART) extends human capabilities by providing researchers with advanced computational and analytical power. As a research tool, AI plays a significant role in literature retrieval and academic writing. In recent years, AI-driven literature search tools, such as Semantic Scholar and Google Scholar, have dramatically enhanced the efficiency with which researchers can access relevant literature. Leveraging natural language processing (NLP) technologies, AI tools can rapidly analyze vast amounts of academic literature, extract key insights, and filter results based on specific research needs \cite{burger2023use, li2019review}. This automated retrieval process not only saves researchers time but also ensures they have access to the most recent and relevant academic findings.

In academic writing, AI tools like ChatGPT and Grammarly provide comprehensive support, from language organization to content generation. These AI tools help researchers in drafting initial paper versions, offering structured text suggestions, and even generating paragraphs appropriate for academic style \cite{somasundaram2023quickly}. Additionally, AI also plays a crucial role in automating reference generation. By integrating with citation management software (e.g., Zotero and Mendeley), AI tools enable researchers to efficiently manage references and automatically format citation lists in various journal styles. This automation not only boosts research productivity but also reduces the likelihood of manual errors, ensuring accuracy in academic writing. 

\subsection{AI as Research Assistant Empowers Human Research}
AI as a research assistant (ARA) extends beyond being a mere tool, taking on more interactive and dynamic roles in the research process. AI assists researchers in experiment design, hypothesis generation, and data collection and analysis, significantly reducing cognitive load and allowing researchers to focus on higher-level conceptual work. 

In brain science field, AI research assistants are increasingly used to monitor and guide experimental setups. They adapt to real-time data and optimize research processes. For example, in neuroimaging studies, AI assistant can suggest optimal scanning protocols and detect anomalies during data collection. This allows researchers to make better decisions during experiments. In brain-computer interfaces (BCI), AI assists in interpreting brain signals, improving communication and understanding \cite{birbaumer2006physiological}. By combining AI with BCI technology, researchers can pinpoint brain areas responsible for movement control. This step is crucial for using brain signals to influence external objects \cite{bell2008control}. In addition, AI-driven deep neural networks also improve brain imaging accuracy, enabling more precise localization of brain regions. Under the guidance of AI assistants, brain scientists can generate precise brain activity images to create functional maps of the brain \cite{darestani2021measuring}.

In social sciences research, AI assistant is essential in transforming large-scale surveys, interviews, and behavioral experiments. Existing social sciences research uses these methods to investigate the social interactions and behavioral patterns of both individuals and groups. However, recent advancements in AI, particularly large language models (LLMs), have the potential to revolutionize the research paradigm of social sciences \cite{grossmann2023ai}. AI assistant can facilitate more dynamic and adaptive research designs by tailoring questionnaires or interview prompts based on real-time feedback from participants \cite{grossmann2023ai}. Furthermore, AI assists in the interpretation of social interaction data, providing preliminary insights that allow researchers to focus on specific patterns or relationships \cite{perez2023serious, farooq2023ai}. For instance, Ziems et al. \shortcite{ziems2024can} analyzed AI's text processing abilities, comparing its performance in replicating hand-coded labels across disciplines such as sociology, political science, psychology, history, literature, and linguistics. The results show that AI is particularly effective in encoding data from political science and sociology. These preliminary studies suggest AI has strong potential as a research assistant in social sciences \cite{korinek2023language}.

\subsection{AI as Research Participant Empowers Human Research}
AI as a research participant (ARP) introduces a fundamentally new paradigm in both brain science and social sciences research. Traditionally, these fields have focused solely on human individuals or groups as research subjects. However, As AI technology continues to develop human-like capabilities and assumes a more prominent role in human life, it should now be recognized as an independent entity actively participating in research. Unlike tools or assistants, which remain subordinate to human direction, AI as a participant actively engages in research process, contributing autonomously and interacting directly with human subjects or researchers.

In brain science, AI systems such as brain-computer interfaces (BCIs) enable real-time interaction with human neural processes. For example, AI-driven BCIs allow for real-time communication between the brain and AI, offering insights into how AI can influence human cognitive functions like attention, memory, and decision-making. A study using fMRI measured human brain activity when subjects interacted with partners including both ``human participants'' and ``AI participants''. When subjects lacked confidence in their decisions, they were more likely to conform to the responses of their partners, either human or AI. This interaction activated the brain's dorsal anterior cingulate cortex (dACC), highlighting how AI influences neural decision-making processes \cite{mahmoodi2022distinct}.

In social sciences, AI as a research participant is even more prominent. AI-driven avatars and conversational agents can interact with human participants in social experiments, simulating social interactions and influencing the dynamics of human behavior. Dell'Aqua et al. investigated the influence of AI on human behavior in multiplayer games, where AI outperformed human players in a cooking task \cite{dell2023super}. The study found that the engagement of AI led to decreased human performance. Besides, the presence of AI in game introduced greater challenges to teamwork and reduce the level of trust among human players. Furthermore, AI's superior performance in group settings often triggered frustration and conflict among human players, while less competent AI can foster human cooperation \cite{traeger2020vulnerable}. These findings suggest that AI's role as research participants can significantly affect group dynamics and social behavior.

Moreover, attitudes toward AI participants can shape human-AI interactions. In another study, individuals interacted with a chatbot that offered varying responses about its cognitive and emotional capacities \cite{pataranutaporn2023influencing}. Those who perceived the AI as having positive intentions viewed it as more reliable and effective. This demonstrates that human perceptions of AI's capabilities and intentions influence the nature of their interactions with it.

\section{Research Methods of Human-AI Joint Research}

\begin{figure*}[h!]
\centering
\includegraphics[width=0.8\textwidth]{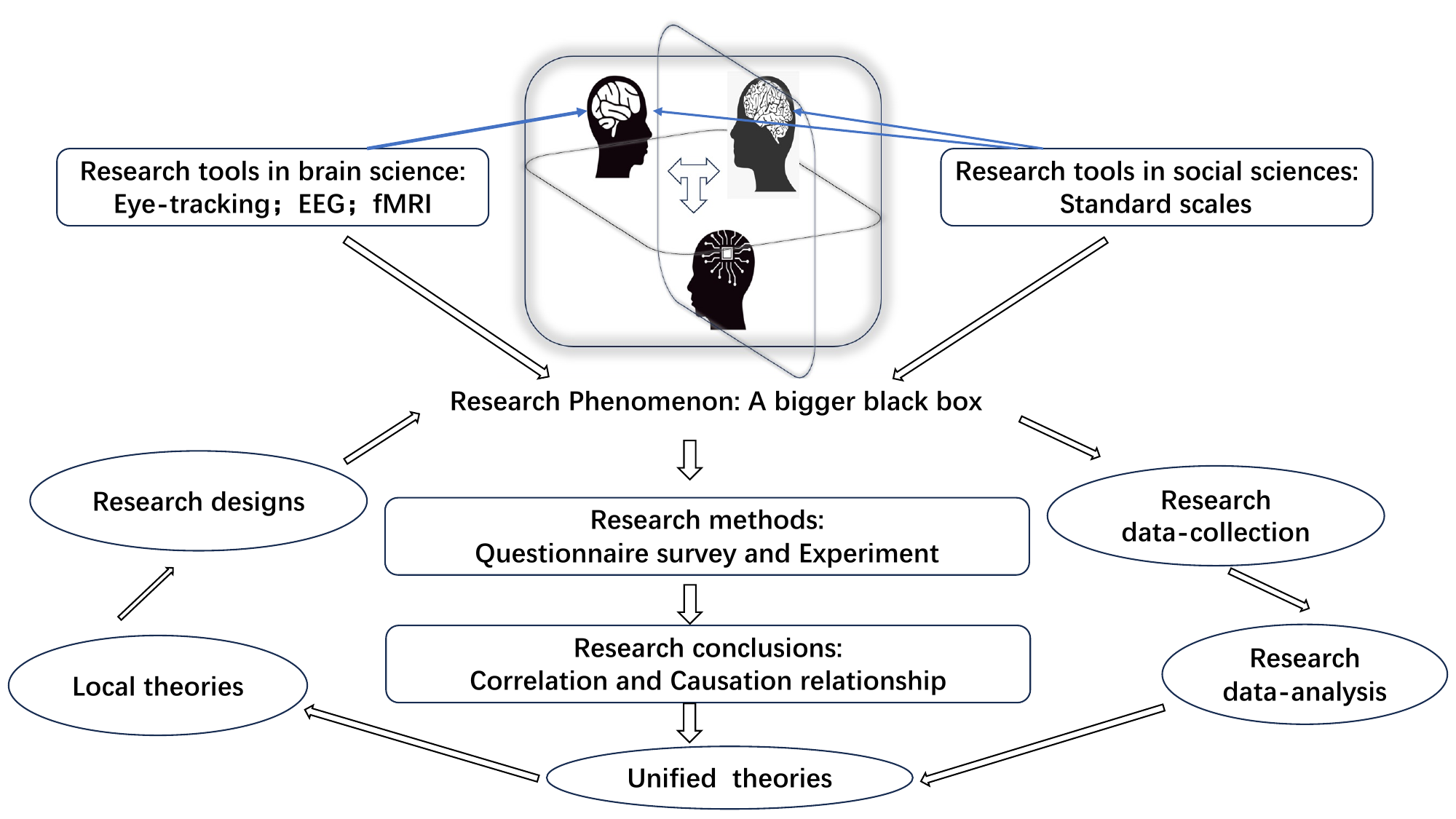} 
\caption{Research methods for conducting human-AI joint research.}
\label{fig:research-method}
\end{figure*}

AI and the human brain respectively serve as black boxes in their respective subject areas. The interaction mechanism between AI and the human brain is a bigger black box. Human-based research methods, such as questionnaires and experiments, may have a positive impact on revealing this research phenomenon. As depicted in Figure \ref{fig:research-method}, researchers in this field can choose either experimental methods or questionnaire methods to collect and analyze research data and finally contribute to forming a unified theory in this field.

In academic research processes, humans need several thinking abilities, such as creativity, critical thinking, and so on. Nowadays, AI technology can comprehend human language, enabling collaboration between humans and AI to complete academic tasks. Therefore, it remains a valuable question how AI affects and reshapes human creativity and critical thinking in academic research processes \cite{xu2023effectiveness}. Exploring human academic abilities involves different research methods in brain science and social sciences, and we will look at many studies to understand this better.

\subsection{Empirical Research of Human-AI Joint Creative Thinking}
Being creative is a fundamental part of being human. It is possible for humans to be more creative by generating new concepts, or to be less creative by staying within their provided knowledge by AI \cite{rafner2023creativity}. One study researched how using AI affects story text creation. A platform was provided for writers to use AI for story inspiration. In this study, participants completed the Divergent Association Task (DAT) to assess their creativity \cite{olson2021naming}. It was a between-subject design in a psychology experiment, including three conditions: one where the story was written only by a person with no help from GenAI, one where a person had one idea from GenAI, and one where a person had five ideas from GenAI. A total of 293 stories were gathered and then evaluated. Access to AI ideas made writers more creative and helped them come up with better stories, especially for writers who are not as creative. However, AI-assisted stories had a higher level of similarity to each other compared to stories created solely by humans \cite{doshi2023generative}.

Another research \cite{cropley2023artificial} focused on the comparison of people and ChatGPT (versions 3 and 4) in their ability to produce innovative ideas. The survey asked people to come up with 10 words that are very different from each other, and their answers were rated based on how different the words were in meaning. ChatGPT came up with more new answers than people did, but the difference was not very big. These findings show that AI could be very useful for people who usually come up with unoriginal ideas.

\subsection{Questionnaire Survey of Human-AI Joint Critical Thinking}
It is widely believed that it is essential for college students to develop critical thinking abilities \cite{daly1998critical}. It involves identifying and solving issues. Critical thinking plays a vital role in helping students learn, analyze information, and make rational choices \cite{alwehaibi2012novel, benitez2013critical}. Educational institutions and businesses consistently emphasize the importance of recognizing critical thinking as a fundamental and necessary skill \cite{roohr2020exploring}. According to studies, first-year college students who are adept at critical thinking and believe they influence their academic achievements are more likely to excel in their classes. It is crucial to recognize the significance of critical thinking in academic achievement \cite{stupnisky2008interrelation}. Moreover, when students rephrase their school readings, it can promote deeper thinking and enable them to meet the elevated cognitive standards of the modern era \cite{li2020cultivation}.

One research examined whether AI can enhance students' critical thinking \cite{jia2024towards}. They conducted a survey and gathered data from 637 college students. Then structural equation modeling was used to test the research hypothesis. The research discovered that the implementation of AI could enhance students' self-assurance and boost their motivation to learn, ultimately leading to an improvement in their critical thinking abilities. This research is a topical questionnaire survey, measuring critical thinking on standard scales and analyzing by structural equation modeling.

\section{Conclusion}
In this paper, we explore the transformative impact of AI on advancing human research, with a particular emphasis on brain science and social sciences. We first review the fundamental aspects of human research, including the human cognition and social interaction. These aspects lay the theoretical foundation for the transition from human research to human-AI joint research. To facilitate this transition, we propose two innovative research paradigms---``AI-Brain Science Research Paradigm'' and ``AI-Social Sciences Research Paradigm''. In these paradigms, we introduce three distinct modes of human-AI collaboration: AI as a research tool (ART), AI as a research assistant (ARA), and AI as a research participant (ARP). In addition, to support these paradigms, we outline the methods for conducting human-AI joint research, incorporating empirical studies on human-AI joint creative thinking and questionnaire surveys on human-AI joint critical thinking. Through a comprehensive review of existing work, we identify the potential benefits and challenges of integrating AI into traditional research paradigms, while advocating for the development of new approaches that effectively integrate both human and AI capabilities. This paper aims to shape future research directions and inspire innovation within this interdisciplinary field, ultimately redefining the collaborative interactions between human researchers and AI.

\appendix

\bibliography{reference}

\end{document}